**Photoexcitation of PbS Nanosheets Leads to Highly Mobile Charge Carriers and Stable Excitons**

*Jannika Lauth,\* Michele Failla, Eugen Klein, Christian Klinke, Sachin Kinge, Laurens D. A. Siebbeles\**

Dr. J. Lauth,
Institute of Physical Chemistry and Electrochemistry, Leibniz Universität Hannover, Callinstr. 3A, D-30167 Hannover, Germany;
Cluster of Excellence PhoenixD (Photonics, Optics, and Engineering – Innovation Across Disciplines), Hannover, Germany
E-mail: jannika.lauth@pci.uni-hannover.de

Dr. J. Lauth, Dr. M. Failla, Prof. L.D.A. Siebbeles
Delft University of Technology, Van der Maasweg 9, NL-2629 HZ Delft, The Netherlands
E-mail: l.d.a.siebbeles@tudelft.nl

E. Klein,
Universität Hamburg, Grindelallee 117, D-20146, Germany
Universität Rostock, Albert-Einstein-Straße 23, D-18059 Rostock, Germany

Prof. C. Klinke
Universität Rostock, Albert-Einstein-Straße 23, D-18059 Rostock, Germany
Swansea University, SA2 8PP, United Kingdom
Universität Hamburg, Grindelallee 117, D-20146, Germany

Dr. S. Kinge
Toyota Motor Europe, Materials Research & Development, B-1930 Zaventem, Belgium

Solution-processable two-dimensional (2D) semiconductors with chemically tunable thickness and associated tunable band gaps are highly promising materials for ultrathin optoelectronics. Here, the properties of free charge carriers and excitons in 2D PbS nanosheets of different thickness are investigated by means of optical pump-terahertz probe spectroscopy. By analyzing the frequency-dependent THz response, a large quantum yield of excitons is found. The scattering time of free charge carriers increases with nanosheet thickness, which is ascribed to reduced effects of surface defects and ligands in thicker nanosheets. The data discussed provide values for the DC mobility in the range 550 - 1000 cm$^2$/Vs for PbS nanosheets with thicknesses ranging from 4 to 16 nm. Results underpin the suitability of colloidal 2D PbS nanosheets for optoelectronic applications.

1. Introduction

Colloidal 2D semiconductor nanosheets (NSs) are interesting for use in optoelectronic devices such as field effect transistors (FETs),[1-2] lasers,[3-4] light-emitting diodes (LEDs)[5] and solar cells.[6-7]. Their band gap can be tuned by varying the NS thickness, while charge carriers and excitons can move efficiently along the lateral dimensions. In addition, colloidal NSs can be cheaply processed from solution. Synthesis methods for 2D PbS-NSs with lateral sizes of several micrometers and a tunable thickness in the range of 3 – 30 nm have been successfully implemented.[2, 8-10] PbS-NSs exhibit more efficient carrier multiplication[11] than PbS quantum dots,[12-14] which is promising for development of high-performance third generation solar cells,[11-17] where multiple electron-hole pairs are generated for each absorbed photon (of sufficient energy). FETs based on individual PbS-NSs show a p-type behavior with a charge carrier mobility of 31 cm$^2$/Vs.[7] By modifying the PbS-NS synthesis with halide ions, a subsequent study reports an n-type behavior of the FETs with a remarkable field-effect mobility of 248 cm$^2$/Vs.[18] Furthermore, measurements of a non-zero circular photo-galvanic effect attributed to Rashba spin-orbit interaction indicates that colloidal 2D PbS-NSs are suitable materials for spintronic devices.[19]

For optoelectronic applications of PbS-NSs, it is important to determine the quantum yields of photogenerated free electrons and holes versus bound electron-hole pairs in the form of excitons (EXs). The exciton binding energy, $E_B$, increases with decreasing NS thickness and dielectric constant of the material surrounding the NSs (the organic ligands).[20-25] Similar to PbS nanocrystals,[26] theoretical calculations report exciton-binding energies in the range of 30 – 70 meV for PbS-NSs with a thickness between 8 and 3 nm.[24] These values are higher than what is expected for bulk PbS (~5-10 meV, determined by the hydrogen atom approach)[22] and underpin the expected increased exciton quantum yield in thinner PbS-NSs.

In this work, the thickness-dependent free charge carrier mobility and the quantum yield of charges versus EXs are investigated by optical pump-terahertz probe spectroscopy (OPTPS). As an all-optical, electrode-less technique, OPTPS is used to determine intrinsic charge carrier transport properties in nanomaterials and avoids difficulties due to contacting of the structures (e.g. different electrodes can affect the determination of the charge mobility).[7, 18] Our results show that the mobility of charge carries in PbS-NSs is thickness-dependent and compares with values reported for bulk PbS. This makes PbS-NSs suitable candidates for next generation ultrathin optoelectronics. In addition, our experimental results can be described by rather high exciton binding energies, as calculated by Yang and Wise.[24]

## 2. Results and Discussion

**Figure 1**(a) shows transmission electron microscopy (TEM) images of the investigated PbS-NSs with different thickness. PbS-NSs were synthesized by injecting 1,1,2-trichloroethane and thioacetamide dissolved in dimethylformamide into a mixture of degassed lead oleate, oleic acid, trioctylphosphine and diphenyl ether, as described previously.[2] Different thicknesses of the PbS-NSs were achieved by varying the temperature and the amount of oleic acid added to the reaction (see also Experimental Section). In line with TEM images and AFM height profile measurements performed in previous studies,[2] a slight variation of the thickness within a single NS thickness is clearly visible. X-Ray diffraction (XRD) patterns of 4, 6 and 16 nm thick PbS-NSs are presented in Figure 1(b). The average thickness of the inorganic PbS-NS only (without the contribution of the oleic acid ligand) is determined by using the Scherrer equation on a Gaussian fit to the FWHM of the prominent (200) reflex at 30° 2Θ in the diffractograms of a drop-casted ensemble of PbS-NSs.[27-28] This reflex represents the <100> direction in the XRD measurements on the substrate. The form factor was selected equal to one due to good

accordance with AFM results for the thickness determination of PbS-NSs in previous studies.[2] Most of the theoretically possible cubic PbS reflexes lack appearance in the diffractograms due to the planar orientation of the NSs on the substrate. The tendency of the NSs to assemble parallel on the substrate can be described as a texture effect, where only lattice planes parallel to the substrate can be measured in the possible angle range. From optical absorption spectra of PbS-NSs in Figure 1(c) and Tauc-plots in the inset, band gaps are determined to be 1.5, 1.4 and 1.1 eV (830, 890 and 1130 nm) for the 4, 6 and 16 nm thick NSs, respectively. The blue-shift with decreasing thickness of the NSs is typical for electronic quantum confinement.

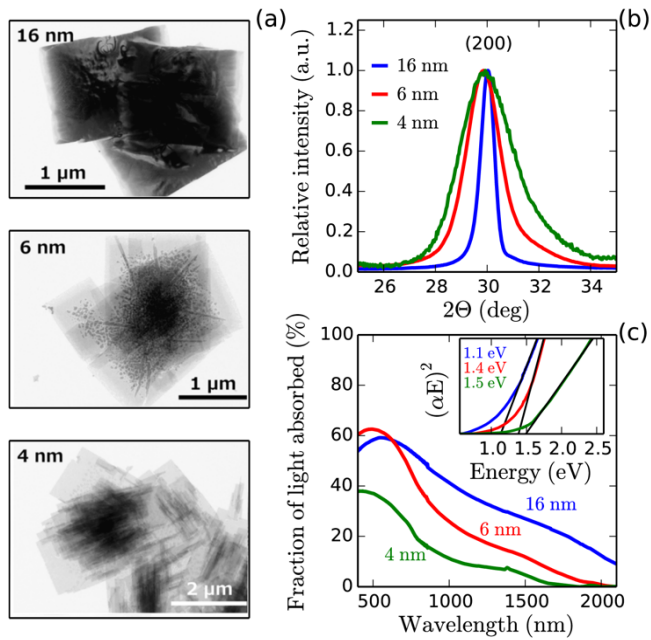

**Figure 1.** (a) TEM images of 4, 6 and 16 nm thick PbS-NSs. (b) XRD patterns of PbS-NSs with different thickness of the inorganic PbS layers as determined by using the Scherrer equation. The peak width of the (200) PbS reflex decreases with increasing sheet thickness and allows for the determination of the inorganic PbS layer without the organic ligands. (c) Optical absorption spectra of PbS-NSs with band gaps between 830 nm (1.5 eV) for the thinnest (4 nm) and 1130 nm (1.1 eV) for the thickest (16 nm) PbS-NSs.

For OPTPS measurements (see also Experimental Section), thin optically dense films of PbS-NSs were drop-casted from a toluene solution onto commercially available quartz substrates (Esco Optics). To prevent degradation, the samples were kept inside a sealed sample-holder under nitrogen during the measurements, and they were stored in a nitrogen-filled glovebox between measurements. The effects of the thickness of PbS-NSs on the dynamics of charge carriers and EXs are investigated by analyzing the transient complex THz signal $S(v,t)$, where $v$ is the frequency of the THz-probe and $t$ is the time delay with the 800 nm optical pump pulse. Within the thin film approximation, $S(v,t)$ is related to the differential transmission of the THz probe, $\Delta E^{THz}(v,t) = E^{THz}_{exc}(v,t) - E^{THz}_0(v)$, where $E^{THz}_{exc}$ and $E^{THz}_0$ are the transmitted THz probe with and without photoexciting the NSs film, respectively, by[29]:

$$S(v,t) = \frac{(1+n_s)c\epsilon_0}{eN_a} \frac{\Delta E^{THz}(v,t)}{E^{THz}_0(v)} \qquad (1)$$

Here, $\sqrt{\epsilon_s} = n_s = 2$ is the refractive index of the quartz substrate, $c$ is the speed of light, $\epsilon_0$ is the vacuum permittivity and $N_a$ is the absorbed photoexcitation density in the NS-films (absorbed photons per unit area). The latter quantity is obtained as $N_a = I_0 F_A$, where $I_0$ is the incident pump laser fluence per unit area and $F_A$ is the fraction of the absorbed photons reported in Figure 1(c). As elaborated in the Experimental Section, $S(v,t)$ is related to the sum of the mobility of free electrons and holes, $\mu_{e,h}(v)$, and the exciton (EX) response, $\mu_{EX}$, weighted by their time-dependent quantum yield, $\phi(t)$, as:

$$S(v,t) = \phi_{e,h}(t)\mu_{e,h}(v) + \phi_{EX}(t)\mu_{EX}(v). \qquad (2)$$

**Figure 2**(a) shows $S(v,t)$ for all PbS-NS samples at a probe frequency $v = 1$ THz and $N_a = 1 \times 10^{14}$ cm$^{-2}$. Filled circles represent the real component, $S_{Re}(t)$, and open circles the imaginary counterpart, $S_{Im}(t)$. $S_{Re}$ clearly decreases with decreasing NS thickness at otherwise similar $N_a$ and pump-probe time $t$, while $S_{Im}$ becomes negative for thinner NSs. Figure 2(b, c) show normalized real and imaginary components of $S(t)$ for $N_a = 1 \times 10^{14}$ and $1 \times 10^{13}$ cm$^{-2}$, respectively. For convenience, the x-axis is reported in logarithmic scale with $t$ shifted by 1 ps. The decay of $S$ with time is related to the decrease of $\phi_i(t)$ due to recombination or trapping at defects. The real and imaginary components both decay similarly and decay times are found to decrease in thinner NSs and at higher $N_a$. For elaborating the thickness dependence of the THz signal, Figure 2(d) shows the frequency dependent $S(\nu)$ for $N_a \sim 5 \times 10^{13}$ cm$^{-2}$ for all samples at $t = 8$ ps and 200 ps after photoexcitation [grey vertical lines in Figure 2(b, c)]. In similar OPTPS experiments performed on GaAs quantum wells (QWs) and Si-bulk at 4 K,[30-31] the time needed to reach thermal equilibrium between excitons and free charge carriers was found to be dependent on the interaction of these species with acoustic and optical phonons, whose population is low at cryogenic temperatures. By pumping the systems well above the bandgap, free carriers were found to convert into EXs on a time scale of hundreds of ps. In the present study, at $T = 300$ K, the charge carrier-phonon interaction is significantly larger so that thermal equilibrium is reached faster. For this reason, the first delay time at 8 ps was chosen to rule out any contribution from hot carriers, and the THz signal is considered to originate from free charge carriers and EXs in thermal equilibrium at the band gap.

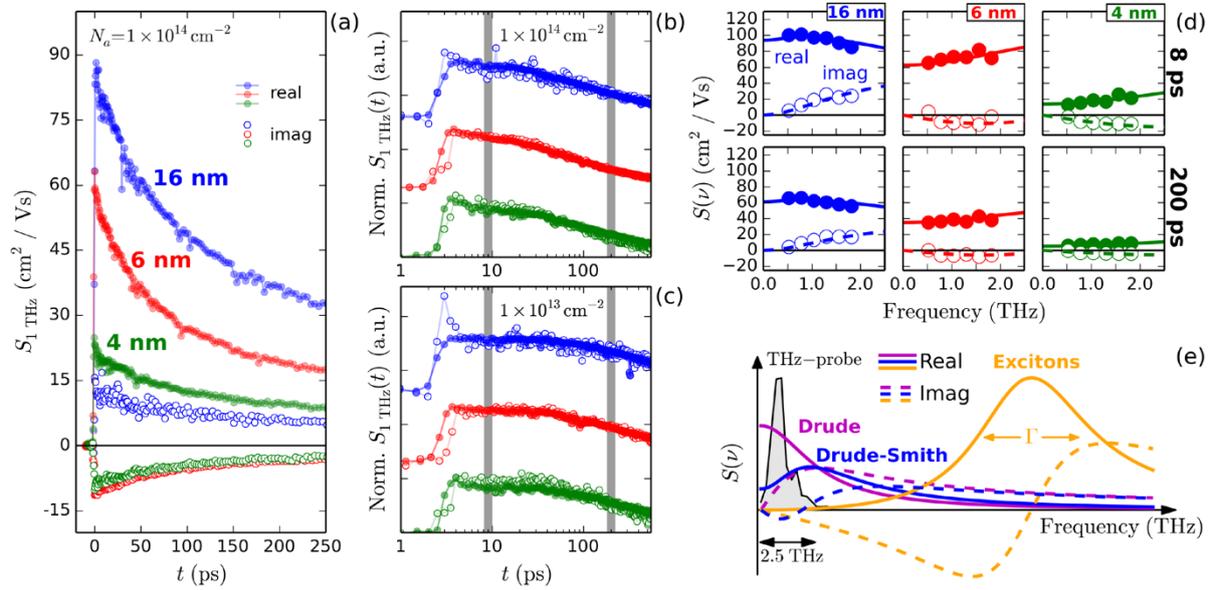

**Figure 2.** (a) Decay-kinetics of the real (filled circles) and imaginary (open circles) component of the transient THz signal $S(\nu = 1$ THz$)$ for 16 (blue), 6 (red) and 4 nm (green) PbS-NSs. $N_a$ indicates the number of absorbed photons per cm$^2$ (see text). (b,c) Comparison between the normalized $S_{Re}$ and $S_{Im}$ at high and low $N_a$, exhibiting the same decay kinetics (d) Transient frequency-dependent THz signals obtained at 8 ps (upper panels) and 200 ps (lower panels)

after excitation [gray vertical lines in (b,c)] at $N_a \sim 5 \times 10^{13}$ cm$^{-2}$. Solid and dashed lines are fits of the real and imaginary components obtained from Equation (2)-(4). (e) Scheme of the response of free charge carriers without backscattering [magenta, Equation (3) with $c = 0$], with backscattering [blue, Equation (3) with $c = -1$] and EXs [orange, Equation (4)].

Figure 2(d) shows that thin films of 6 and 4 nm thick PbS-NSs reveal increasing $S_{Re}$ values and negative, decreasing, $S_{Im}$ values with frequency, both at 8 and 200 ps after photoexcitation. In stark contrast, for 16 nm thick PbS-NSs, $S_{Re}$ decreases with frequency, while its imaginary counterpart $S_{Im}$ remains positive. The THz signal due to free charge carriers is proportional to the sum of the electron and hole mobility. In order to take into account any possible scattering of free carriers with structural defects, we describe their response according to the Drude-Smith like behavior [32-33]:

$$\mu_{e,h} = \mu_e + \mu_h = \frac{e\tau}{m^*} \frac{1}{1-i\omega\tau}\left(1 + \frac{c}{1-i\omega\tau}\right) \qquad (3)$$

Here, $m^* = m_e m_h/(m_e + m_h)$ is the reduced effective mass obtained from effective masses of electrons ($m_e$) and holes ($m_h$), $\tau$ is the scattering time, which is assumed to be the same for electrons and holes, $c$ is the backscattering parameter and $\omega = 2\pi\nu$ is the radial THz frequency. The real and imaginary mobility obtained from Equation (3) are shown in Figure 2(e). The real mobility, obtained without considering backscattering (solid magenta curve obtained with $c = 0$), decreases with frequency, while the imaginary mobility (dashed magenta curve) first increases with frequency, reaches a maximum at $\omega\tau = 1$, and decreases subsequently. The blue curve shows the maximum effect of backscattering by taking $c = -1$, i.e a backscattering angle of 180 deg. The imaginary component is lowered towards negative values and the maximum of the real component is shifted from $\omega = 0$ towards higher frequencies.

The EX response originates from transitions from the lowest exciton state ($n = 0$, and binding energy $E_B = E_0$) to higher states ($n > 0$, with binding energy $E_n$). This is described by[30]:

$$\mu_{EX}(\omega) = \frac{e}{im^*}\sum_n \frac{f_{B,n}\omega}{\omega_{B,n}^2-\omega^2-i\omega\Gamma} \quad (4)$$

where $\omega_{B,n} = 2\pi\nu_{B,n} = |E_B - E_n|/\hbar$ and $\Gamma$ is the transition broadening, which is considered to be independent of $n$. Factors $f_{B,n}$ in Equation (4) are oscillator strengths which, for 2D-systems, can be analytically calculated as[30]:

$$f_{Bn} = \frac{2m^*a_B^2}{\hbar^2}(E_n - E_B)\frac{\left(n+\frac{1}{2}\right)^5 n^{2n-3}}{(n+1)^{2n+5}} \quad (5)$$

where $a_B = 4\pi\hbar^2\epsilon_0\epsilon/e^2m^*$ is the Bohr radius, $\epsilon_0$ is the vacuum permittivity and $\epsilon$ is the dielectric function experienced by EXs within the material. $\epsilon$ depends on the dielectric function of the PbS-NSs and the lower dielectric function of the surrounding medium (oleic acid ligands). The real component of Equation (4) is shown as the orange solid curve in Figure 2(e). This describes the absorption of the THz probe, which is maximum at resonance for which $\omega = \omega_{B,n}$. In the example of Figure 2(e), this resonance is assumed to occur at much higher frequency than the probing THz field (frequencies between 0 and 2.5 THz, gray area), and to have a narrow linewidth $\Gamma$. The imaginary counterpart (dashed orange curve) describes the EX polarizability, $\alpha = -e\,Im(\mu_{EX})/\omega$. Values of $Im(\mu_{EX})$ are negative at low frequency, decrease approximately in a linear fashion and change sign at the frequency corresponding to the first EX resonance. Values of $E_B$ and $\Gamma$ determine how the THz signal is affected by EXs.

Thickness dependent effective masses of electrons and holes in Equation (3) and (4) were obtained from the energy-dispersion of the conduction and valence bands, which were calculated by the k·p method, analogous to the work of Yang and Wise.[24] **Figure 3**(a) shows $m_e, m_h$ and $m^*$, as obtained by fitting the calculated bands with a parabolic function (note: since the bands deviate from a parabolic dispersion, substantially different values reported in Ref. [24] arise from the smaller k-range chosen in our fitting procedure). It is seen that the effective masses decrease with thickness.

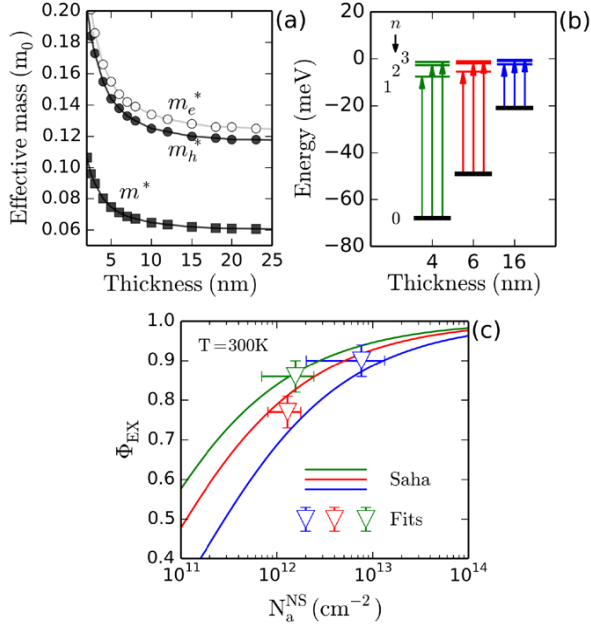

**Figure 3.** (a) Thickness-dependence of the effective mass for electrons ($m_e$), holes ($m_h$) and related reduced effective mass ($m^*$) obtained from k·p calculations (see text). (b) Calculated excitonic states and relative transition energies (vertical colour arrows) for different PbS-NS thicknesses. Thick black lines are exciton-binding energies from Ref. [24]. (c) $\Phi_{EX}$ obtained from the Saha model (color lines) as a function of the absorbed photons per unit area in a single NS, $N_a^{NS}$, and from fits in Figure 2(d) (triangles).

Calculated values of $m^*$ for PbS-NSs studied here are reported in **Table 1**. By using values of $E_B$ for PbS-NSs with thickness of 4 and 6 nm from Yang and Wise,[24] and by assuming the value for 16 nm NSs to be ~21 meV [i.e. between the value for 8 nm NSs and the bulk value, see Equation S1 in the Supporting Information], we calculate the energies of higher exciton states according to $E_n = \frac{E_B}{4\left(n+\frac{1}{2}\right)^2}$.[25] Energies obtained in this way are given in Table 1 and shown in Figure 3(b). The remaining unknown parameters in Equation (3), (4) and (5) are $c, \tau, a_B$ and $\Gamma_{EX}$. These parameters are obtained by fitting Equation (2) to the experimental data in Figure 2(d) together with the quantum yield $\phi_{e,h}$ and $\phi_{EX}$. For photoexcitation densities considered in Figure 2, the decay of the THz response is negligible on a time scale up to 10 ps, and hence we assume $\phi_{e,h} + \phi_{EX} = 1$.

**Table 1.** $E_B$ values for PbS-NSs with different thickness as calculated in Ref. [24]. Reduced effective mass $m^*$ calculated by means of the four-band effective mass model, as presented in [24]. Free charge carrier scattering time ($\tau_{e,h}$), backscattering parameter ($c$) and quantum yield ($\Phi_{e,h}$), EX quantum yield ($\Phi_{EX}$), excitonic transition broadening ($\Gamma$), Bohr radius ($a_B$) obtained from fits presented in Figure 2 (c) at 8 ps.

| PbS-NS [nm] | $m^*$ [$m_0$] | $E_B$ [meV] | $\tau_{e,h}$ [fs] | $c$ | $\Phi_{e,h}$ | $\Phi_{EX}$ | $\Gamma$ [meV] | $a_B$ [nm] |
|---|---|---|---|---|---|---|---|---|
| 4 | 0.08 ± 0.1 | 68 | 25 ± 4 | -0.82 ± 0.05 | 0.14 ± 0.04 | 0.86 ± 0.04 | 153 ± 30 | 3.7 ± 0.5 |
| 6 | 0.07 ± 0.1 | 49 | 29 ± 4 | -0.62 ± 0.05 | 0.23 ± 0.04 | 0.77 ± 0.04 | 140 ± 75 | 4.5 ± 1.0 |
| 16 | 0.06 ± 0.1 | 21 | 33 ± 4 | -0.0 ± 0.1 | 0.1 ± 0.04 | 0.9 ± 0.04 | 131 ± 80 | 6.0 ± 1.0 |

Fits as described above reproduce well the real and imaginary components of the THz signal at 8 ps (see solid and dashed colour lines in Figure 2(d), respectively). Values of the abovementioned adjustable parameters are reported in Table 1. It shows that photoexcitation predominantly leads to the formation of EXs and the quantum yield of free charge carriers is found to range between 0.1 and 0.2. The response of the two species is compared in Figure S1 of the Supporting Information showing that the EX response is rather weak due to small $a_B$ values [see Equation (5)], as discussed later in the text. In Figure S3 of the Supporting Information, we attempt to fit the data by considering the response as only due to free charge carriers ($\Phi_{e,h} = 1$), with the mobility related to the Drude-Smith model. In this case, the measurement data can only be reproduced with values for effective masses well above those calculated here and in Ref. [24]. Moreover, the Drude-Smith model fails to describe the frequency dependence of the imaginary component for 6 nm NSs. These two observations exclude the photoexcitation of free charge carriers only and confirm that the formation of EXs must be taken into account in order to explain the data.

To prove the reliability of values found for $\Phi_{EX}$, we consider the Saha model for 2D systems,

which describes the ratio between photogenerated free carriers and EXs in equilibrium.[30] As stated before, since the decay kinetics remain constant in the first 10 ps of the measurement for all samples, by means of the Saha model, $\Phi_{EX}$ is calculated as a function of the number of absorbed photons per unit area in a single PbS-NS, $N_a^{NS}$ (see Supporting Information), and by using $E_B$ and $m^*$ values reported in Table 1. Calculations are shown in Figure 3(c) by solid lines. Triangles are EX fractions obtained from fits shown in Figure 2(d) at 8 ps. By taking into consideration the uncertainty in the fitted values of $\Phi_{EX}$ (vertical error bars) and the uncertainty of the calculated $N_a^{NS}$ (horizontal error bars), found values agree well with what predicted by the Saha model, confirming the validity of our approach.

From fits shown in Figure 2(d), we find that the scattering time of free charges, $\tau$, increases with the PbS-NS thickness, which is ascribed to a smaller probability of scattering on surface defects or ligands in the case of thicker NSs. The same argument can be used to describe the thickness trend for $c$, which decreases with increasing NS thickness. In Figure S3, a comparison between normalized spectra at 8 and 200 ps shows that the frequency-response remains identical for all samples. This implies that the THz signal decays according to a similar decrease of $\Phi_{e,h}$ and $\Phi_{EX}$, therefore accounting for the similar decay kinetics of the imaginary and real components in Figure 2(b) and (c). Indeed, the response at 200 ps can be fit by considering the same scaling factor for $\Phi_{e,h}$ and $\Phi_{EX}$ only, while other parameters remain constant. At 200 ps, we find a 35, 44 and 62 % reduction of both $\Phi_{e,h}$ and $\Phi_{EX}$ for the 16, 6 and 4 nm thick PbS-NS samples, respectively. We infer a stronger reduction of both fractions in thinner NSs as due to enhanced trapping or recombination processes.

From values of $m^*$ and $\tau$ in Table 1, the sum of the DC mobility of electrons and holes can be calculated as $\mu_{DC} = e\tau/m^*$. This yields values of $550 \pm 100$, $700 \pm 100$ and $1000 \pm 150$ cm$^2$/Vs, for 4, 6 and 16 nm thick PbS-NSs, respectively. These values are comparable with those from Hall measurements on bulk PbS ($\mu_{e,h} \simeq 1000$ cm$^2$/Vs),[34] and surpass values reported for PbS-

NSs FETs, where the evaluation of the mobility could be affected by the device contacts.[18]

Values for the bandwidth of the EX transition $\Gamma$ reported in Table 1 are similar to those reported for the excitonic response in monolayer transition metal dichalcogenides.[35-36] The substantial uncertainty arises from the fact that EX transitions are well above the frequency range of our THz probe. Bohr radii obtained for 4 and 6 nm thick NSs agree well with the distance between electrons and holes calculated in Ref. [24]. The response for the 16 nm thick NSs, by taking into account the EX contribution, could only be reproduced with $a_B = 6$ nm, which is unreasonable since this value is comparable to the value of the thinner NSs. For 16 nm NSs, we expect a value for $a_B$ between 5.5 nm (8 nm thick NSs, Ref.[24]) and the PbS bulk value (~18 nm[37]). The small value for the 16 nm thick NSs could be due to the fact that the above stated $E_B$ is overestimated for this thickness. In addition, Equation (5) for the oscillator strength is valid for 2D-systems, and it may be no longer applicable for thicker NSs. From the definition of $a_B$ given above and by taking average values for the thinnest samples, $a_B \simeq 4.0$ nm and $m^* \simeq 0.075\ m_0$, we obtain $\epsilon^{\text{PbS-NSs}} \simeq 6.0$. This value is reasonable, since it is close to average of the high-frequency dielectric constant of PbS, $\epsilon_\infty^{\text{PbS}} = 17$,[38] and that of the OA ligands, $\epsilon_\infty^{\text{OA}} \sim 2$.

We infer that the description of the Coulomb interaction in an exciton requires the use of an effective dielectric constant given by the contribution of the high-frequency dielectric constant of the material where they are confined (rather than the static one, $\epsilon_0^{\text{PbS}} = 175$[38]), and the dielectric constant of the surrounding medium. Our finding agrees with the use of $\epsilon_\infty^{\text{PbS}}$ in calculations on excitons in PbS-NSs by Yang and Wise,[24] where the effective dielectric constant results from the contribution of $\epsilon_\infty^{\text{PbS}}$ and $\epsilon_\infty^{\text{OA}}$. In OPTPS studies on GaAs quantum wells by Kaindl et al.,[30] the static dielectric constant was used to describe the properties of excitons. However, for GaAs, it is challenging to distinguish between the high-frequency and static dielectric constant because of their similar values ($\epsilon_\infty^{\text{GaAs}} = 10.9$, $\epsilon_{st}^{\text{GaAs}} = 12.9$).[39]

## 3. Conclusion

In conclusion, we have determined the frequency-dependent complex THz-response of photogenerated free charge carriers and excitons in colloidal PbS-NSs with different thickness. From analysis of our data, we find DC mobilities as high as 550 to 1000 cm$^2$/Vs for 4 to 16 nm thick PbS-NSs, rendering them leading-edge thin film 2D materials. The increase of the mobility is due to an enhanced scattering time of charge carriers in thicker NSs, which is likely due to the reduced influence of interaction with surface defects and ligands. Our data agree with substantial exciton binding energies from previously reported theoretical calculations. The present study emphasizes the excellent potential of colloidal semiconductor NSs for innovative optoelectronics.

## 4. Experimental Section

*Synthesis and characterization of 2D PbS-Nanosheets*

The method described by Bielewicz *et. al.* was used to synthesize 2D PbS-NSs.[2] The use of halogenated solvent additives and a large lead to sulfur ratio in comparison to a regular PbS nanocrystal synthesis, yields 2D PbS-NSs with lateral dimensions of several micrometers (see TEM images in Figure 1). The thickness of the PbS-NSs (4, 6 and 16 nm) is tuned by varying the amount of oleic acid used in the synthesis. TEM images are obtained with a JEOL-JEM 1200 operating at 100 kV. The thickness of the PbS-NS films is evaluated with a Dektak Stylus profiler to be 300 ± 100 nm, 440 ± 70 nm and 140 ± 60 nm for 4, 6 and 16 nm thick PbS-NS samples, respectively.

*Absorption spectra*

The fraction of light absorbed by the PbS-NS samplesd is obtained with a Perkin Elmer Lambda 1050 spectrometer equipped with an integrating sphere, which allows for the correction of the reflection by the samples and the quartz substrate.

*Optical pump - terahertz probe spectroscopy (OPTPS)*

A Coherent Libra laser system is used to create a 1.4 kHz pulse train at 800 nm. The output of the Libra consists of two beams, one of which is uncompressed. The other, compressed beam, with a pulse duration of ~60 fs, is led to a beam splitter: 50% of the laser fluence is used to optically pump the sample, the other 50% is used to generate THz radiation in an 0.2 mm thick ZnTe crystal, which is subsequently focused on the sample. The optical pump fluence is varied by using neutral density filters. Care is taken that the diameter of the optical pump beam (~2 mm in diameter) is always larger than the focused THz probe (~1.5 mm) to ensure homogeneous excitation. The detection of the transmitted THz probe is performed in a single-shot detection method in which the entire THz waveform is probed at once. This is done using a chirped optical pulse, which is created by varying the degree of pulse compression of the uncompressed Libra output via an external compressor. This chirped pulse is overlapped in another 0.5 mm thick ZnTe crystal with the transmitted THz probe to detect its electric field by electro-optic sampling. After transmission through the ZnTe crystal, the chirped pulse is dispersed with a grating onto a set of CCD array detectors to obtain the entire THz waveform. Both, the THz generation beam as well as the optical pump beam, are led over delay stages, making it possible to vary either the THz time delay $t$ or the pump delay $\tau$ with respect to the fixed THz detection pulse. A scan of the THz time delay is necessary (once) to relate the different pixels of the CCD array to different times.

By measuring the pump-induced differential transmission of the terahertz (THz) waveform $\Delta E(t_p, t) = E_{\text{exc}}(t_p, t) - E_0(t_p)$, OPTPS allows the determination of the complex, frequency-dependent photoconductivity $\Delta\sigma(\nu, t)$,[32-33, 40-43] where $\nu$ is the frequency of the THz-probe. Here, $E_0(t_p)$ and $E_{\text{exc}}(t_p, t)$ are the transmitted THz pulses before and after photoexcitation, respectively. The time $t_p$ is the detection time of the THz probe-pulse, while $t$ is the time-delay between the THz-probe and the 800 nm pump-pulse. Within the thin-film approximation, the transient photoconductivity of NS films with thickness $L$ can be evaluated as:[32-33, 43]

$$\Delta\sigma(\nu,t) = \frac{(1+n_s)\epsilon_0}{L}\frac{\Delta E(\nu,t)}{E_{\text{exc}}(\nu)} \qquad (6)$$

where $\sqrt{\epsilon_s} = n_s = 2$ is the refractive index of the quartz substrate, $c$ is the speed of light and $\epsilon_0$ is the vacuum permittivity. Accessing $\Delta\sigma$ via OPTPS can allow to distinguish between the contribution of different photogenerated species to $\Delta\sigma$, such as free charge carriers, excitons (EXs), trions, plasmons or polaritons.[32-33, 43] OPTPS has been extensively used to determine the dielectric properties of 2D NSs,[29, 35, 44-46] NCs[47-49] and organic polymers.[50-51] The transient conductivity, $\Delta\sigma(\nu,t)$ is related to the sum of the frequency-dependent mobility of free unbound electron-hole (e-h) pairs ($\mu_{e,h}$) and EXs ($\mu_{EX}$). These are weighted by their time-dependent density $n_{e,h}(t)$ and $n_{EX}(t)$, according to:

$$\Delta\sigma(\nu,t) = e[n_{e,h}(t)\mu_{e,h}(\nu,t) + n_{EX}(t)\mu_{EX}(\nu,t)], \qquad (7)$$

where $e$ is the elementary charge. By determining the absorbed photoexcitation density $N_a = I_0 F_A$ (absorbed photons per unit area) - where $I_0$ is the incident pump laser fluence per unit area and $F_A$ is the fraction of the absorbed photons reported in Figure 1(c) - one finds $n_i(t) = \Phi_i(t)N_a/L$.

Here, $\Phi_i = n_i/N_a$ is the quantum yield of the $i$-th species such that $\Phi_{e,h} + \Phi_{EX} = 1$ at $t = 0$. In this work, the thickness dependence of the optoelectronic properties of thin PbS-NS films are investigated by analyzing the transient THz signal, $S(\nu,t)$, which does not require the explicit determination of the thickness of PbS-NSs films. By substituting Equation (7) into (6), and considering the relation between the carrier density and quantum yield above, we evaluate the product of the quantum yield and response of photogenerated EXs and free charge carriers as:[29]

$$S(\nu,t) = \sum_i \Phi_i(t)\mu_i(t) = \frac{(1+n_s)c\epsilon_0}{eN_a}\frac{\Delta E(\nu,t)}{E_{\text{exc}}(\nu)} \qquad (8)$$

**Supporting Information**
Supporting Information is available from the Wiley Online Library or from the author.

**Acknowledgements**

Dr. J. Lauth and Dr. M. Failla contributed equally to this work.

We thank Dr. Michiel Aerts and Dr. Juleon M. Schins for contributing to an early stage of the data analysis and manuscript. This work is part of the research program TOP-grants with project number 715.016.002, which is financed by the Netherlands Organization for Scientific Research (NWO). C.K. gratefully acknowledges financial support of the European Research Council via the ERC Starting Grant "2D-SYNETRA" (Seventh Framework Program FP7, Project: 304980) and the German Research Foundation DFG for financial support in the frame the Heisenberg scholarship KL 1453/9-2. E.K. and C.K. thank the German Research Foundation DFG for financial support in the frame of the Cluster of Excellence "Center of ultrafast imaging CUI".
J.L. acknowledges funding by the Deutsche Forschungsgemeinschaft (DFG, German Research Foundation) under Germany's Excellence Strategy within the Cluster of Excellence PhoenixD (EXC 2122, Project ID 390833453).

Received: ((will be filled in by the editorial staff))
Revised: ((will be filled in by the editorial staff))
Published online: ((will be filled in by the editorial staff))
16

Time-resolved terahertz spectroscopy is applied to probe stable excitons and highly mobile charge carriers (up to 1000 cm$^2$/Vs) in colloidal 2D PbS nanosheets of different thickness, rendering the materials highly promising for optoelectronics.

**Thin-film optoelectronics**

*Jannika Lauth,\* Michele Failla, Eugen Klein, Christian Klinke, Sachin Kinge, Laurens D. A. Siebbeles\**

**Photoexcitation of PbS Nanosheets Leads to Highly Mobile Charge Carriers and Stable Excitons**

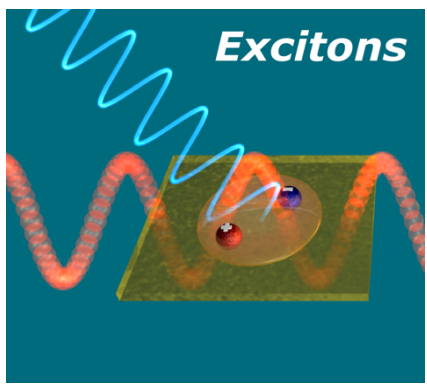

Supporting Information

**Photoexcitation of PbS Nanosheets Leads to Highly Mobile Charge Carriers and Stable Excitons**

*Jannika Lauth,\* Michele Failla, Eugen Klein, Christian Klinke, Sachin Kinge, Laurens D. A. Siebbeles\**

1. Evaluation of the exciton binding energy for 16 nm PbS-NSs

For 16 nm PbS-NSs, we obtain a value $E_B$ = 21 meV, by taking it as the average value between the value calculated by Yang and Wise for 8 nm thick PbS-NSs (40 meV) and the bulk value.[24] The latter is obtained from the relation:

$$E_B = 13.6\ eV \cdot \frac{m^*}{m_0} \frac{1}{\epsilon_\infty} \qquad (S1)$$

where $m^*$ is the reduced effective mass of the exciton given by $(m_e m_h)/(m_e + m_h)$, $\epsilon_\infty$ is the high frequency dielectric constant and $m_0$ is the free electron mass.[25] From Equation (S1), the binding energy is $E_B$ = 2.7 meV by considering $m_e$ = 0.12, $m_h$ = 0.11 and $\epsilon_\infty$ = 17.[26]

2. Number of absorbed photons per NS

The number of absorbed photons per unit area in a single nanosheet is calculated as:

$$N_a^{NS} = \frac{N_a}{n_{NS}},$$

where $N_a$ is the absorbed photoexcitation density in the NS-film (absorbed photons per unit area) and $n_{NS}$ is the number of NS within the film, which is evaluated by knowing the film thickness (see Experimental Section) and the thickness of a single NS.

3. Fits of the THz response in the frequency domain

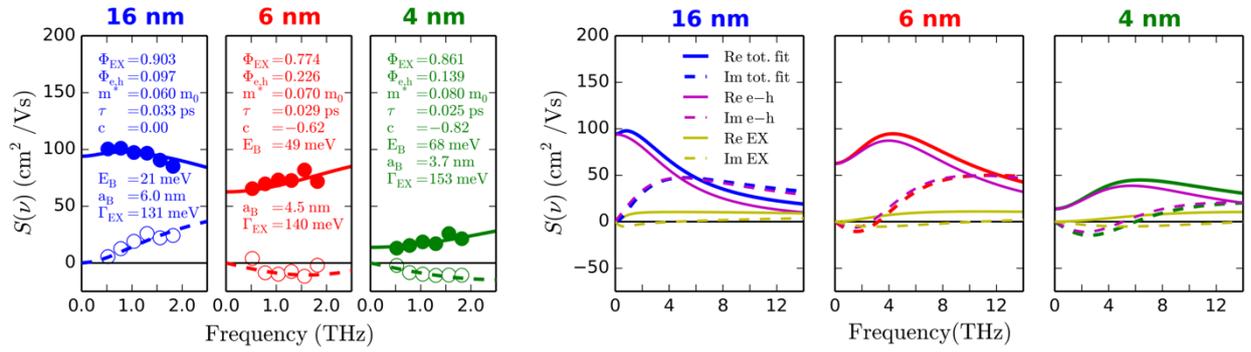

**Figure S1.** Fits of the THz response at 8 ps for PbS-NSs with different thickness (indicated at the top of panels). In the right panels the total response (same color as the sample), EX response (yellow) and free charge carrier response (magenta) from fits are shown in a wider frequency range. Note that with obtained $a_B$ values the response of more than 70 % of photogenerated charge carriers at $t$ = 8 ps is weak.

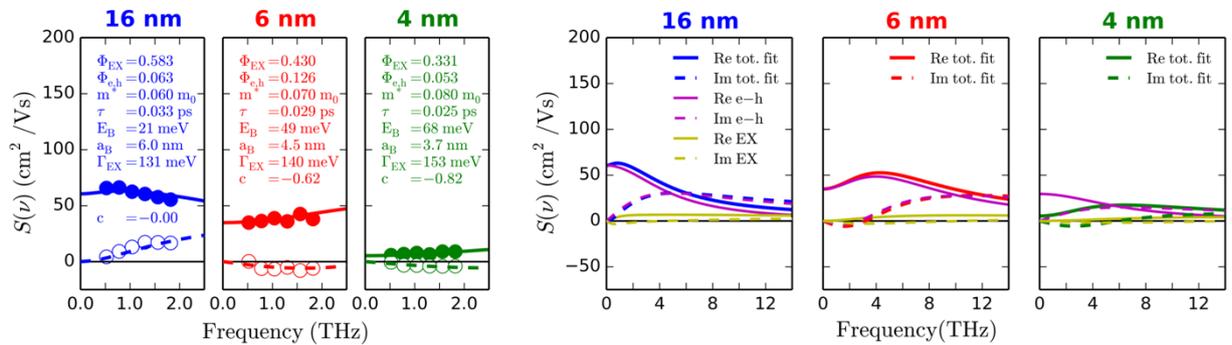

**Figure S2.** Fits of the THz response at 200 ps for PbS-NSs with different thickness (indicated at the top of panels). In the right panels the total response (same color as the sample), EX response (yellow) and free charge carrier response (magenta) from fits are shown in a wider frequency range.

4. Normalized THs response at 8 and 200 ps

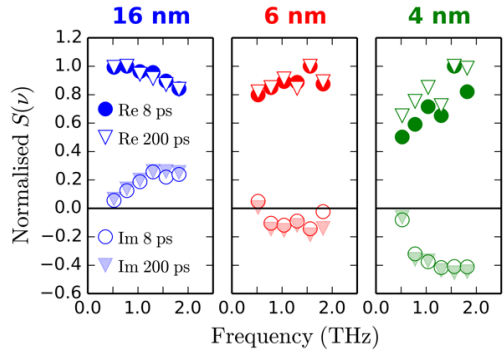

**Figure S3.** Comparison of normalized THz spectra at 8 and 200 ps after photoexcitation, showing that the later time response can be described by a scaling factor. In the main manuscript, this is related to the reduction of the fraction of free charge carriers only.

5. Fits of the THz response by considering free charge carriers only

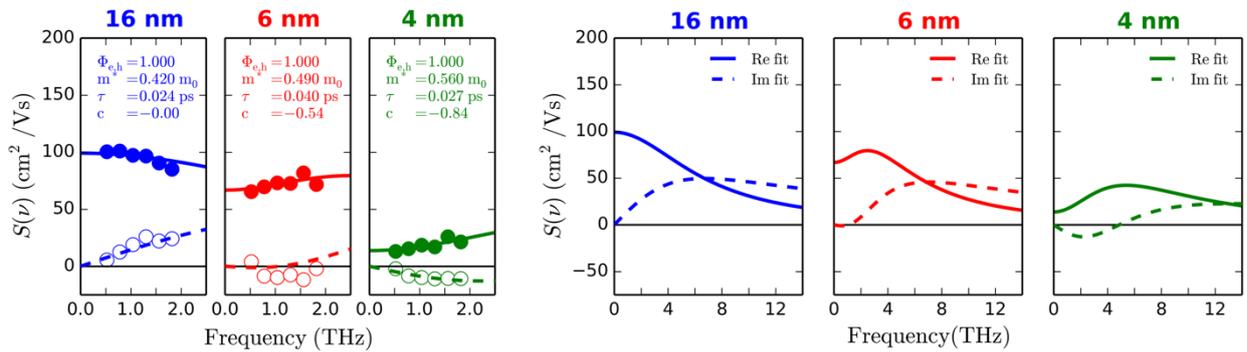

**Figure S4.** An attempt for fitting the measured data by considering free charge carriers only and the Drude-Smith model. The response is well reproduced for the 16 and 4 nm samples, while a noticeable disagreement is apparent for the imaginary component of 6 nm thick PbS-NSs. Obtained effective masses deviate substantially from values reported in the main manuscript.